\documentclass[showpacs,amsmath,amssymb,aps,10pt,reprint,superscriptaddress,prb]{revtex4-1}
\usepackage{graphicx}
\usepackage{bm}
\usepackage[breaklinks=true,colorlinks=true,linkcolor=blue,urlcolor=blue,citecolor=blue]{hyperref}
\usepackage{graphics}
\usepackage{dcolumn}
\usepackage{amsmath,amssymb}
\usepackage{mathdots}
\usepackage{natbib}
\usepackage{soul,color}
\usepackage{epstopdf}
\usepackage[note-name]{notes2bib}
\begin{document}
\title{Room-temperature annealing effects on the basal-plane resistivity of optimally doped YBa$_2$Cu$_3$O$_{7-\delta}$ single crystals}

\author{G. Ya. Khadzhai}
\author{R. V. Vovk}
\author{N. R. Vovk}
\author{S. N. Kamchatnaya}
    \affiliation{Physics Department, V. Karazin Kharkiv National University, 61077 Kharkiv, Ukraine}
\author{O.~V.~Dobrovolskiy}
    \affiliation{Physikalisches Institut, Goethe University, 60438 Frankfurt am Main, Germany}
    \affiliation{Physics Department, V. Karazin Kharkiv National University, 61077 Kharkiv, Ukraine}

\begin{abstract}
We reveal that the temperature dependence of the basal-plane normal-state electrical resistance of optimally doped YBa$_2$Cu$_3$O$_{7-\delta}$ single crystals can be with great accuracy approximated within the framework of the model of s-d electron-phonon scattering. This requires taking into account the fluctuation conductivity whose contribution exponentially increases with decreasing temperature and decreases with an increase of oxygen deficiency. Room-temperature annealing improves the sample and, thus, increases the superconducting transition temperature. The temperature of the 2D-3D crossover decreases during annealing.
\end{abstract}
\maketitle

\section{Introduction}

Long-term stability of the electrical transport characteristics of modern multipurpose materials based on high-$T_c$ superconducting cuprates
is one of the most crucial requirements for their use in instruments and devices. Thus far, one of the most asked-for compounds is the so-called 1-2-3 system RBa$_2$Cu$_3$O$_{7-\delta}$ (where R = Y or other rare earths) \cite{Wum87prl,Sol16prb}. This is stipulated by several important factors. Firstly, this compounds have rather high critical characteristics, namely the superconducting transition temperature $T_c> 90$\,K, noticeably above the nitrogen liquefaction temperature. Secondly, one can rather easily vary both, the superconducting and normal-state characteristics of the system by complete or a partial substitution of its constituents or changing the deviation degree from the oxygen stoichiometry \cite{Gin89boo}. Finally, there are well-established technologies for the fabrication of cast, thin-film and single-crystal samples of rather big sizes, the latter being especially favorable for fundamental investigations.

At the same time, the presence of a labile component (oxygen) in the compound leads to the appearance of a nonequilibrium state in a particular sample. The latter can easily be induced by the application of a high pressure \cite{Sad00prb}, an abrupt temperature change \cite{Vov14ltp}, or ensue in the course of a long-term storage or exploitation (aging) \cite{Mar95apl}. Thus far, rather specific diffusion processes take place in the system \cite{Kla98pcs} that, in turn, contributes to a phase segregation \cite{Bey89nat}, structural relaxation, and the appearance of various superstructures. All these factors noticeably affect the charge and heat transfer mechanisms in the system, as well as the appearance of peculiar electrical transport phenomena such as fluctuation conductivity \cite{Asl68pla}, pseudogap anomaly \cite{Sad05prb}, incoherent electronic transport \cite{And91prl} and so on. According to contemporary views, it is these nontrivial phenomena in the normal state which are expected to be the key to understanding the microscopic nature of high-$T_c$ superconductivity which remains unresolved so far, despite a more than 30-year-long history of intensive experimental and theoretical investigations.

The electrical properties of superconducting cuprates in the normal state are known to differ not much from those of ordinary metals, see e.g. Ref. \cite{Mak00ufn}. It was shown that the experimental dependences of the basal-plane electrical resistance of underdoped YBa$_2$Cu$_3$O$_{7-\delta}$ single crystals in the normal state, $\rho_{nab}(T)$, can be described well by the Bloch-Gr\"uneisen formula \cite{Mak00ufn} accounting for scaterring of the conduction electrons on phonons and defects. In this case the dependence $d\rho_{nab}(T)/dT$ exhibits a smeared maximum at $T_m \approx 0.35\theta$, where $\theta$ is the Debye temperature amounting to $\theta \simeq 500$\,K for underdoped samples. Accordingly, $\rho_{ab}(T)$ has a characteristic inflection peculiar to phonon scattering. At $T > \theta$ the Bloch-Gr\"uneisen expression asymptotically tends to a straight line with increasing temperature, while with a decrease of the temperature it turns down from the high-temperature extrapolation $\rho \propto T$, that is associated with a transition from elastic to inelastic phonon scattering.

For optimally doped YBa$_2$Cu$_3$O$_{7-\delta}$ single crystals there is no maximum in $d\rho_{nab}(T)/dT$ that might be stipulated by a decrease of $T_m$ to $T_m \leq T_c$, that is by a decrease of $\theta$ to $\theta \leq 200$\,K. It is worth noting that (i) phonon scattering always takes place and (ii) the resistivity values of YBa$_2$Cu$_3$O$_{7-\delta}$ single crystals correspond to those of metal systems with a pseudogap, such as amorphous alloys, quasicrystals, transition metal dichalcogenides and so on \cite{Mot90boo}. A downturn of  $\rho_{nab}(T)$ from $\rho \propto T$ may also be caused by an excess conductivity which has an exponential increase with decreasing temperature and is associated with the pseudogap in YBa$_2$Cu$_3$O$_{7-\delta}$.

Thus, in order to approximate the dependence $\rho_{ab}(T)$ of YBa$_2$Cu$_3$O$_{7-\delta}$ single crystals in a wide temperature range it is necessary to take into account the various mechanisms of the conductivity and charge carriers scattering. In the present work, we study the effect of annealing at room temperature on the approximation parameters and on the superconducting characteristics of the samples --- the point which has insufficiently been investigated in the literature so far. This makes it possible to clarify the physical nature of the individual conduction and scattering mechanisms, as well as the ways to affect them.

%In particular, we fit the experimental curves $\rho_{ab}(T)$ of optimally doped YBa$_2$Cu$_3$O$_{7-\delta}$ single crystals to the Bloch-Gr\"uneisen relation with an account for the fluctuation conductivity. We also analyze the room-temperature annealing effects on the fitting parameters and the superconducting characteristics of the samples.

\section{Experimental}
The YBa$_2$Cu$_3$O$_{7-\delta}$ single crystals were grown by the solution-melt technique in a gold crucible as in Ref. \cite{Vov12jms}. As is known \cite{Vov12jms}, a tetra-ortho structural transition takes place in YBa$_2$Cu$_3$O$_{7-\delta}$ upon saturation with oxygen. This transition leads to a crystal twinning thus minimizing its elastic energy. To obtain twin-free samples, crystals were untwinned in a crucible at a pressure of $30-40$\,GPa at $420^\circ$C as described in Ref. \cite{Gia89ltp}. To obtain a controllable homogeneous distribution of oxygen, the samples were further annealed in an oxygen atmosphere at $420^\circ$C for 7 days.

For resistive measurements several crystals were selected. Electrical contacts were created in the standard 4-probe geometry by applying a silver paint on the crystal surface. This followed by attachment of silver conductors with $0.05$\,mm in diameter and a three-hour-long annealing at $200^\circ$C in ambient atmosphere. This procedure has allowed us to obtain a transient contact resistance of less than $1\,\Omega$ and to conduct resistance measurements at transport currents up to $10$\,mA in the $ab$-plane. The measurements were done in the temperature-sweep mode. Temperature was measured using a platinum resistor thermometer. The superconducting transition temperature was determined at the point of maxima in the dependences $d\rho_{ab}(T)/dT$ in the region of the superconducting transition.

To reduce the oxygen content, the samples were annealed in an oxygen flow at $620^\circ$C for two days. After annealing the samples were quenched to room temperature within $2-3$ minutes, mounted on the holder, and cooled down to liquid nitrogen temperatures within 10-15 minutes. All measurements were done while warming the samples up. For investigations of the effect of room-temperature annealing, after the first measurement of $\rho(T)$, the samples were kept at room temperature for 20 hours and repetitive measurements were performed. The final series of measurements was done after a room-temperature annealing of the samples for 3 days. In what follows we discuss the data acquired on one typical sample.

\section{Results and discussion}
\subsection{Normal resistance and excess conductivity}
\begin{figure}
    \centering
    \includegraphics[width=0.5\textwidth]{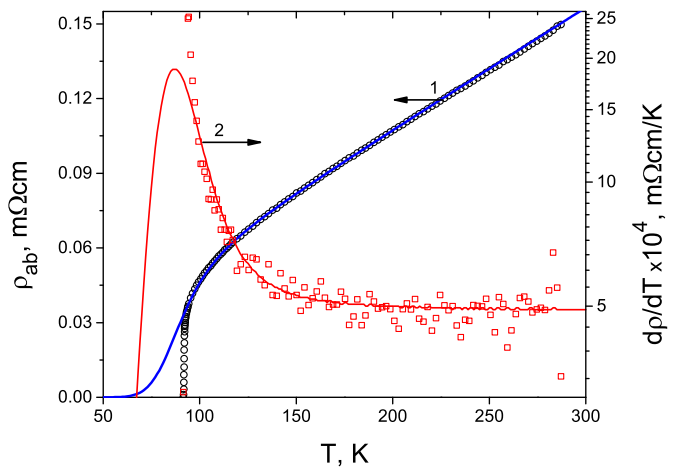}
    \caption{Temperature dependences of the electrical resistivity $\rho_{ab}(T)$ of the optimally doped YBa$_2$Cu$_3$O$_{7-\delta}$ single crystal after quenching from $620^\circ$C. 1 -- $\rho_{ab}(T)$. Symbols: experiment. Solid line: Calculation by Eqs. \eqref{e1} and \eqref{e2}. 2 - $d\rho_{nab}(T)/dT$. Symbols: Evaluated experimental data. Solid line: calculation by Eqs. \eqref{e1} and \eqref{e2}.}
    \label{f1}
\end{figure}
Figure \ref{f1} displays the experimental temperature dependence of the basal-plane electrical resistivity, $\rho_{ab}$, (symbols) of the optimally doped YBa$_2$Cu$_3$O$_{7-\delta}$ single crystal after quenching from $620^\circ$C. The curves measured after different stages of annealing are qualitatively similar. The dependence $\rho_{ab}(T)$ has been revealed to fit (solid line in Fig. \ref{f1}) the Bloch-Gr\"uneisen formula
\begin{equation}
\label{e1}
    \rho_{app}(T) = [\rho^{-1}_n(T)  +  b_0 (\exp^{T_1/T} -1)]^{-1},
\end{equation}
where
\begin{equation}
\label{e2}
    \rho_n(T) = \rho_0 + C_3\left(\frac{T}{\theta}\right)^3 \int_0 ^{\theta/T} \frac{e^x x^3 dx}{(e^x - 1)^2},
\end{equation}
and $\rho_0$ is the residual resistivity. The fitting parameters in Eqs. \eqref{e1} and \eqref{e2} were determined by minimums of the error least squares, which does not exceed 1\%. The fitting parameters are reported in Table \ref{table}.

Figure \ref{f1} also displays the temperature dependences of the derivative $d\rho_{nab}(T)/dT$ (symbols) and the temperature derivative of Eq. \eqref{e1} (solid line). Due to the presence of the exponential term in Eq. \eqref{e1} the maximum in $d\rho_{app}(T)/dT$ ensues at $T_m \approx 87$\,K, while the maximum in $d\rho_n(T)/dT$ at $T_{mBG} \approx 50$\,K.
\begin{table}[b!]
\begin{tabular}{|c|c|c|c|}
\hline
                & Quenching             & Annealing         & Anealing\\
                & from $620^\circ$C     & for $20$\,hours   & for $92$\,hours\\
                \hline
$\rho_0$, m$\Omega$cm & 0.0126          & 0.0119            & 0.0140\\
$C_3$, m$\Omega$cm    & 0.141           & 0.134             & 0.134\\
$\theta$, K           & 145             & 138               & 142\\
$T_1 = U_{pg}/k$, K   & 1060            & 1060              & 1070\\
$b_0$, (m$\Omega$cm)$^{-1}$  & $1.05\times10^{-4}$ &$1.06\times10^{-4}$& $9.9\times10^{-5}$\\
$T_c$, K           & 91.88             & 91.95               & 92.01\\
$\Delta T_{c0.5}$, K & 0.366          & 0.325               & 0.233\\
$T_{cross}$, K        & 92.09          & 92.05               & $\geq$92.01\\
$\xi_c(T_{cross})$, \AA     & 0.28     & 0.25                & ---\\
\hline
\end{tabular}
   \caption{Fitting parameters for the basal-plane electrical resistivity of the optimally doped YBa$_2$Cu$_3$O$_{7-\delta}$ single crystal by Eqs. \eqref{e1}, \eqref{e2}, and \eqref{e3}.}
   \label{table}
\end{table}

We note that at $T \gg \theta$ it is $\rho_n \approx \rho_0 + \left(\frac{C_3}{2\theta}\right)T$ which yields the mentioned linear temperature dependence $\rho_{ab}(T)$ at high temperatures, namely at $T \geq 170$\,K in our case.

\begin{figure}[t!]
    \centering
    \includegraphics[width=0.95\linewidth]{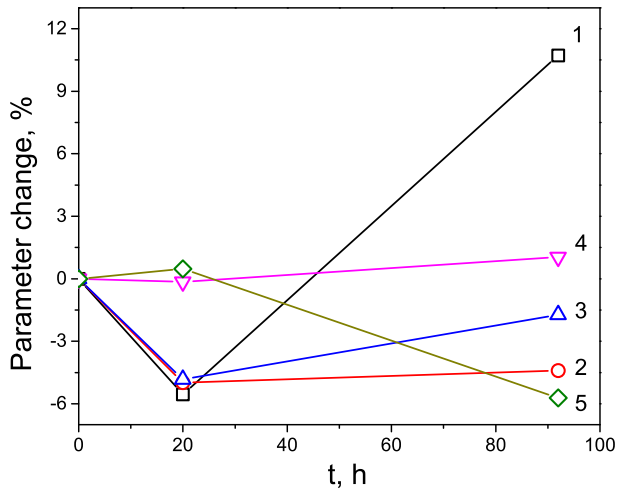}
    \caption{Relative changes of the fitting parameters by Eqs. \eqref{e1} and \eqref{e2} in dependence of the annealing time at $20^\circ$C. 1 -- $\rho_0$, 2 -- $C_3$, 3 -- $\theta$, 4 -- $T_1$, 5 -- $b_0$.}
    \label{f2}
\end{figure}

The relative changes of the fitting parameters for Eqs. \eqref{e1} and \eqref{e2} in dependence on the time of annealing at $20^\circ$C are presented in Fig. \ref{f2}. One sees that for the optimally doped YBa$_2$Cu$_3$O$_{7-\delta}$ single crystal the 92\,hour-long room-temperature annealing does not lead to significant changes in the parameters of charge scattering on phonons. Namely, the phonon scattering coefficient $C_3$ and the Debye temperature $\theta$ only exhibit a weak tendency to increase. We note that the Debye temperature for optimally doped single crystals is much smaller than for underdoped ones in a reference measurement. Since the crystal lattice of YBa$_2$Cu$_3$O$_{7-\delta}$ is easily deformed under the relative shift of the layers, and $\theta \propto a^{-1}$ ($a$ is the interatomic distance), the small values of $\theta$ obtained by approximating the temperature dependence of the resistance of a perfect single crystal can be due to the preferential charge carriers scattering on the vibrations of atoms along the $c$ axis ($\theta \propto d^{-1}$, $d$ is the distance between the layers). The values of $\theta$ deduced by us agree well with the results  of Ref. \cite{Mak00ufn} where the temperature dependence of the heat capacity of YBa$_2$Cu$_3$O$_{7-\delta}$ was described by a model that takes into account the anisotropy of the crystal lattice of this material. An increase of the Debye temperature for underdoped samples in this case can be related to to the isotropization of the phonon spectrum due to an increase of the number of defects --- oxygen vacancies.

The residual resistivity associated with charge carriers scattering on defects also demonstrates a tendency to increase. This attests to an increase of the number of defects, first of all, of oxygen vacancies. Along with the increase of the superconducting transition temperature $T_c$, please refer to Table \ref{table}, this indicates a tendency of the concentration of oxygen vacancies to some saturation value corresponding to the optimal $\delta$, and in turn, to maximal $T_c$.

The term describing the paraconductivity is characterized by the increasing temperature $T_1$, whereas the coefficient $b_1$ exhibits a tendency to decrease, see Fig. \ref{f2} and Table \ref{table}. Taking into account the behavior of the residual resistivity as well as the fact that in underdoped YBa$_2$Cu$_3$O$_{7-\delta}$ single crystals the excess conductivity is not observed in the background of phonon scattering \cite{Mak00ufn}, one may assume a suppression of the paraconductivity by defects: With an increase of $\delta$ the manifestation of the excess conductivity, represented by the coefficient $\beta$, is weakened.

We note that metal systems with a pseudogap can exhibit a temperature dependence described by the Bloch-Gr\"uneisen expression. The pseudogap value chiefly depends on the material composition \cite{Mot90boo}, while its temperature dependence is weak and it is primarily determined by the thermal expansion of the material. A certain role in this may be played by specific mechanisms of quasiparticle scattering \cite{Apa02prb,Ada94ltp,Cur11prb} stipulated by the presence of a structural and kinematic anisotropy in the system.

For this reason the pseudogap feature associated with temperature can also appear in the case when the pseudogap is a precursor of the superconducting transition. In this case the pseudogap transits into the superconducting gap with a decrease of temperature \cite{Lar09boo}. Accordingly, the second term in Eq. \eqref{e1} can be considered as a contribution to the conductivity of occasionally appeared Cooper pairs with a coupling energy of $kT_1 \sim 0.1$\,eV.\\

\subsection{Superconducting transition}
\begin{figure}
    \centering
    \includegraphics[width=0.85\linewidth]{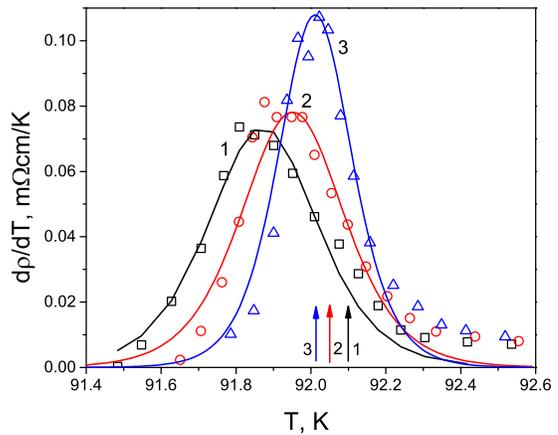}
    \caption{Temperature derivatives of the resistivity, $d\rho(T)/dT$, in the vicinity of the superconducting transition for different annealing times. 1 -- original curve after quenching from $620^\circ$, refer to Fig. \ref{f1}; 2 -- after annealing at $20^\circ$C for 20 hours; 3 -- after annealing at $20^\circ$C for 92 hours. Symbols: evaluated experimental data. Solid lines: calculations by Eq. \eqref{e3}. Vertical arrows indicate the respective crossover temperatures.}
    \label{f3}
\end{figure}

In Fig. \ref{f1} one sees that at $T<95$\,K Eqs. \eqref{e1} and \eqref{e2} do not properly describe the dependence $\rho_{ab}(T)$ as the superconducting transition sets on. This transition occurs in some temperature range stipulated by the degree of sample inhomogeneity. It can be described as a smeared superconducting transition with an inclusion function describing $\rho_{sc}(T)$ as a stepped function smeared near $T_c$ \cite{Rol83boo}. The derivative of the inclusion function is a curve with a finite maximum
\begin{equation}
    \label{e3}
    \frac{d\rho_{sc}(T)}{dT} = \frac{\rho_{sc} e^{-Z}}{w(1 + e^{-Z})^2},\qquad Z = \frac{T- T_c}{w}.
\end{equation}

The maximum in $d\rho_{sc}(T)/dT$ is attained at $T = T_c$. The parameter $w$ characterizes the transition smearing: The full width at half height of the maximum amounts to $\Delta T_{c0.5}\approx 3.5w$. For the high-$T_c$ 1-2-3 system the validity of this description was corroborated in Refs. \cite{Kha17fnt}.

Figure \ref{f3} displays the resistivity temperature derivatives $d\rho_{sc}(T)/dT$ in the vicinity of the superconducting transition for different annealing times. The fitting parameters to Eq. \eqref{e3} are reported in Table \ref{table}. One sees that in the course of annealing $T_c$ increases (the maxima in $d\rho_{sc}(T)/dT$ are shifted to the right), while their width associated with the inhomogeneity of the superconducting state is decreasing. In this way the improvement of the structural perfectness contributes to the increase of $T_c$.

\begin{figure}
    \centering
    \includegraphics[width=0.85\linewidth]{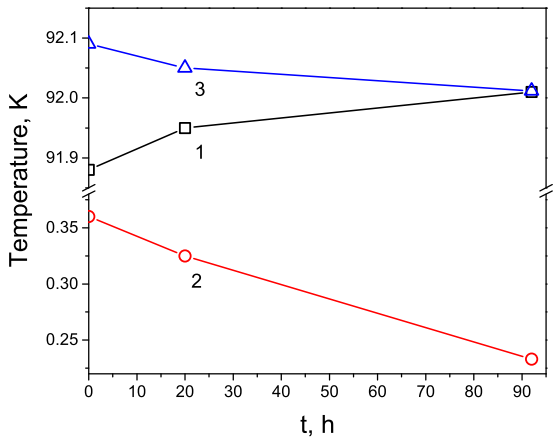}
    \caption{Parameters of the superconducting state for different annealing times. 1 -- $T_c$; 2 -- $\Delta T_{c0.5}$; 3 -- $T_{cross}$.}
    \label{f4}
\end{figure}

The parameters of the superconducting state in dependence on the annealing time are presented in Fig. \ref{f4}. One sees that in the course of annealing $T_c$ is increasing (curve 1), while its width $\Delta T_{c0.5}$ is decreasing (curve 2). Hence, reduction of the crystal imperfection leads to the maximal $T_c$ (optimal $\delta$ value).

In Ref. \cite{Lar09boo} for the basal-plane conductivity near $T_c$ the following expression is reported
\begin{equation}
\label{e4}
    \sigma^{AL}_{ab} = \frac{e^2}{16 \hbar d} \frac{1}{\sqrt{\varepsilon(\varepsilon + r)}},
\end{equation}
where $d = 11.7$\,\AA~is the interlayer distance \cite{Keb89prb}, $\varepsilon = \frac{T-T_c}{T_c}\ll 1$, $r = (4(_\downarrow c^\uparrow 2(0)))/d^\uparrow 2$.
Formula \eqref{e4} describes a 2D-3D crossover occurring in a certain temperature range. Namely, for $\varepsilon \ll r$ it yields $\sigma_{ab}^{AL}\propto(\varepsilon r)^{-1/2}$ (3D mode). For $\varepsilon \gg r$ one has $\sigma_{ab}^{AL}\propto (\varepsilon)^{-1}$ (2D mode). The condition $\varepsilon_{cross} = r$ determines the conventional point of the crossover.

\begin{figure}
    \centering
    \includegraphics[width=0.4\textwidth]{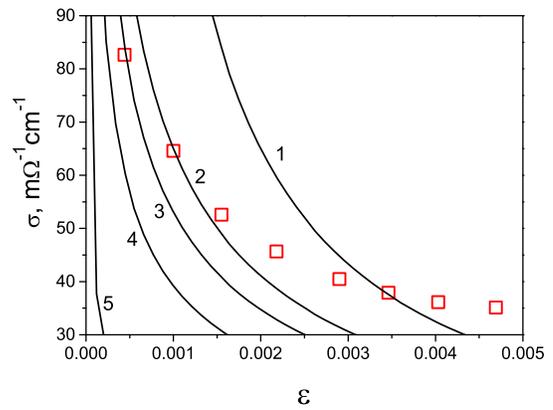}
    \caption{Dependences $\sigma(\varepsilon)$ in the region of the superconducting transition. Lines are the curves $\sigma_{ab}^{AL} = f(\varepsilon)$ calculated by Eq. (\ref{e4}) for $r =0$ (1), 0.001 (2), 0.002 (3), 0.005 (4), and 0.01 (5). Symbols are the experimental data $\sigma(\varepsilon) = [\rho(\varepsilon)]^{-1}$ in the region of the superconducting transition for in Fig. \ref{f1}.}
    \label{f5}
\end{figure}

Figure \ref{f5} displays the dependences $\sigma_{ab}^{AL} = f(\varepsilon)$ for a series of values of $r$ in comparison with the experimental data $\sigma(\varepsilon) \approx [\rho(\varepsilon)]^{-1}$ in the region of the superconducting transition for the sample in the original state (Fig. \ref{f1}). One sees that the experimental data for $\varepsilon \leq 0.005$ fall between $r = 0$ (curve 1) and $r = 0.005$ (curve 3). Each experimental value of $\sigma(\varepsilon)$ corresponds to a certain value of $r$ which decreases with increasing $\varepsilon$, that is with an increase of the temperature. One also sees that for these experimental points $\varepsilon \approx r$ that is they are in the crossover region.

We assume that for $\varepsilon \leq 0.05$ the experimental values of the resistivity are only determined by the transition into the superconducting state $(\rho_{sc})^{-1} \approx \sigma^{AL}_{ab}$ and determine $r$ by Eq. \eqref{e4} in this temperature range. It turns out that this parameter is not constant, but it rather decreases with increasing temperature so that, we believe, one can write $r = (4(_\downarrow c^\uparrow 2(T)))/d^\uparrow 2$. Then, the condition for the crossover is an intersection of the curves $r(T_{cross}) = \varepsilon(T_{cross})$. These intersections are observed for the original curve and the first curve measured after annealing. For the third curve there is no intersection, therefore we assume $T_{cross3} \geq T_{c3}$. The values of $T_{cross~i}$ thus deduced are reported in Table \ref{table} and are indicated by the arrows in Fig. \ref{f3}. Finally, given the values of $r_{cross}$ and $T_{cross}$ one can estimate the respective coherence length $\xi_c(T_{cross} = 0.5d (r_{cross})^{1/2}$. These values are also reported in Table \ref{table} and they are very close to the data of Ref. \cite{Fri89prb}.

\section{Conclusion}
Our results attest to that (i) The temperature dependence of the basal-plane electrical resistance (above $T_c$) of optimally doped YBa$_2$Cu$_3$O$_{7-\delta}$ single crystals can be with a great accuracy approximated within the framework of the model of s-d electron-phonon scattering with account for the fluctuation conductivity whose contribution exponentially increases with decreasing temperature. (ii) The paraconductivity term decreases with increasing oxygen deficiency. (iii) Annealing stipulates an improvement of the sample imperfectness and, in this way, an increase of $T_c$. (iv) The 2D-3D crossover temperature is shifted in the same direction as $T_c$ does in the course of annealing.

\section*{Acknowledgements}
The research leading to these results has received funding from the European Union's Horizon 2020 research and innovation program under Marie Sklodowska-Curie Grant Agreement No. 644348 (MagIC).

%\bibliographystyle{unsrt}

% \bibliographystyle{elsarticle-num}
% \bibliographystyle{elsarticle-harv}
% \bibliographystyle{elsarticle-num-names}
% \bibliographystyle{model1a-num-names}
% \bibliographystyle{model1b-num-names}
% \bibliographystyle{model1c-num-names}
% \bibliographystyle{model1-num-names}
% \bibliographystyle{model2-names}
% \bibliographystyle{model3a-num-names}
% \bibliographystyle{model3-num-names}
% \bibliographystyle{model4-names}
% \bibliographystyle{model5-names}
% \bibliographystyle{model6-num-names}

%\bibliography{D:/bibliobase/ybco}
%\balance

\end{document}